\documentclass[aps,pra,twocolumn,superscriptaddress,showpacs,longbibliography]{revtex4-1}
\usepackage{bm} 
\usepackage{graphicx}
\usepackage{epsfig}
\usepackage{hyperref}
\usepackage{natbib}
\usepackage{amsmath}
\usepackage{amssymb}
\usepackage{color}

\newcommand{\rmi}{{\rm i}}
\newcommand{\rmd}{{\rm d}}
\newcommand {\e}{{\rm e}}

\renewcommand {\i}{{\rm i}}
\renewcommand {\Re}{\mathop{\mathrm{Re}}\nolimits}
\renewcommand {\Im}{\mathop{\mathrm{Im}}\nolimits}

\begin{document}

\title{Phase spectroscopy of topological invariants in photonic crystals}

\author{A.\,V.\,Poshakinskiy}\email{poshakinskiy@mail.ioffe.ru}
\author{A.\,N.\,Poddubny}
\affiliation{Ioffe Institute, 
St.~Petersburg 194021, Russia}
\author{M.~Hafezi}
\affiliation{Joint Quantum Institute, NIST/University of Maryland, College Park, Maryland 20742, USA}
\affiliation{Department of Electrical Engineering and Institute for Research in Electronics and Applied Physics, University of Maryland, College Park, MD 20742, USA}

\begin{abstract}
We propose a method of measuring topological invariants of a photonic crystal through phase spectroscopy. We show how the Chern numbers can be deduced from the winding numbers of the reflection coefficient phase. An explicit proof of existence of edge states in system with nonzero reflection phase winding number is given. The method is illustrated for one- and two-dimensional photonic crystals of nontrivial topology.
\end{abstract}

\pacs{78.67.Pt, 03.65.Vf, 73.43.Cd}









\maketitle


\section{Introduction}

The presence or absence of topologically protected states at the edge of a material is determined by the topology of its bulk Bloch states~\cite{Bernevig_book}. This topology can be characterized by  integer invariants,  that determine whether the material is topologically trivial or non-trivial. Thus, development of methods to measure the topological invariants is  one of the most important problems in the field.
The topological invariants of the electron gas in external magnetic field (Chern numbers) are directly related to the quantized Hall conductivity~\cite{TKNN}. Determining the topological invariants in various recently proposed  counterparts of this system~\cite{Lu2014} is though more complicated.
Particularly, the time-of-flight images were demonstrated to contain information about topological invariants of a cold-atom system~\cite{Alba2011,Zhao2011,Goldman2013}. Theoretically predicted possibility to extract Zak phase, Chern numbers or $\mathbb{Z}_2$ topological invariants from semiclassical dynamics of a wave packet~\cite{Price2012,Wang2013,Abanin2013,Liu2013,Grusdt2014,Ozawa2014} was recently experimentally realized for cold atoms in one-dimensional optical lattice~\cite{Atala2013}. 
Photonic systems are more preferable for realization of different  measurement schemes  due to easier optical access to microscopic properties as compared to conventional electronic topological insulators or cold-atom systems.
The methods to measure the topological numbers by tracing the individual edge states fingerprints in transmission spectra~\cite{Hafezi2014} or by manipulating the single unit cell and directly measuring the Bloch function~\cite{Bardyn2013arXiv} were proposed for a lattice of coupled ring waveguides. 
The winding number of the scattering matrix eigenvalues was shown to determine the number of edge states and topological invariants~\cite{Meidan2011,Fulga2012,Rudner2013,Pasek2014,Hu2014arXiv}, although no clear way to measure them has been proposed yet. 
Recently, the relation between the surface impedance and Zak phase for a centrosymmetric one-dimensional photonic crystal was revealed~\cite{Xiao2014}. 
In this work we show that the topological indexes of a photonic system can be measured via the phase of the reflection coefficient, that is accessible in simplest optical experiments. We consider both  one-dimensional (1D) Aubry-Andr\'e-Harper (AAH) photonic crystal~\cite{aubrey,Lang2012,dassarma2013,Poshakinskiy2014prl} and  two-dimensional (2D) lattice of tunneling-coupled resonators with synthetic magnetic field~\cite{Hafezi2011,Hafezi2013}. 
The direct correspondence between (i) the winding numbers of the reflection coefficient, (ii) the Chern numbers, and (iii) the presence of the edge states is proven. This provides the general recipe to access the edge states and topological indexes from outside the structure by optical means.

\section{Topological indexes in 1D photonic crystal.} 
First, we explain the proposed method for the 1D photonic crystal inspired by the AAH model~\cite{aubrey,Poshakinskiy2014prl}. We consider the stack of alternating layers $\mathcal A$ and $\mathcal B$, see Fig.~1a, where all the layers $\mathcal A$
have the same width $d_{\mathcal A}$, while the width of the layers $\mathcal B$ is periodically modulated
as~\cite{aubrey}
\begin{equation}\label{eq:pos}
d_{\mathcal B,n} = \bar d_{\mathcal B} [ 1 + \eta \cos \left( 2 \pi b n + \varkappa \right) ]\:.
\end{equation}
Here, $\bar d_{\mathcal B}$ is the  width of unmodulated layers, $\eta$, $b$, and $\varkappa$ are the modulation strength, frequency and phase; we focus on the rational frequency case $b=P/Q$, where $P$ and $Q$ are the integers with no common factor. 
The difference between the  dielectric constants of the layers $\varepsilon_{\mathcal A}$ and $\varepsilon_{\mathcal B}$ leads to the Bragg reflection of the normally propagating light. Namely, in the absence of modulation the system forms  an elementary photonic crystal with the period $d=d_{\mathcal A}+\bar d_{\mathcal B}$ and the Bragg gaps in the energy spectrum around the integer multiples of $\omega_B = \pi c/ ({\sqrt{\varepsilon_{\mathcal A}}} d_{\mathcal A} + {\sqrt{\varepsilon_{\mathcal B}}} \bar d_{\mathcal B})$. The modulation ($\eta\ne 0$) enlarges the unit cell size and the period from
 $d$ to  $D=Qd$, and drastically modifies the spectrum by splitting each Bragg gap into $Q$ gaps.
  Importantly, the system exhibits cyclic evolution when the modulation phase $\varkappa$  is continuously changed by $2\pi$. In the real system, it might be possible to realize this continuous variation if the modulation is induced by a running acoustic wave. This cyclic behavior  allows one to map the 1D system to 2D one, and  introduce the topological indices --- Chern numbers~\cite{Lang2012,dassarma2013}.

The considered 1D photonic system is completely characterized by the dependence of the dielectric function $\varepsilon_\varkappa(z)$ on the coordinate along the growth axis $z$, and on the external parameter $\varkappa$. In the case of normal light incidence the wave equation  at the frequency $\omega$ for the  electric field component $E(z)$ perpendicular to $z$ axis reads
\begin{equation}\label{Max0}
\frac{\rmd^2}{\rmd z^2}\, E(z) + \frac{ \omega^2}{c^2}\,\varepsilon_\varkappa(z)\, E(z) = 0\:.
\end{equation}
The  properties of the  eigenstates of Eq.~\eqref{Max0} can be most conveniently
analyzed in the reciprocal space.
Performing the spatial Fourier transform we obtain the system of coupled equations for the harmonics $E_{K-G}$ with the  wave vectors $K-G$,
\begin{equation}\label{Max}
(q^2-K^2)E_K + q^2\sum\limits_G \epsilon_{G,\varkappa} E_{K-G} = 0\:,
\end{equation}
where $G=2\pi g/D$ $(g=0,\pm 1,...)$ is the reciprocal lattice vector, $q=\omega \sqrt{\varepsilon_{0,\varkappa}}/c$, $\varepsilon_{G,\varkappa} = (1/D) \int_0^D \rmd z\, \varepsilon_\varkappa(z) \,\e^{-\rmi G z} $, $\epsilon_{G,\varkappa} = \varepsilon_{G,\varkappa}/\varepsilon_{0,\varkappa}$. In the case $d_\mathcal{A} \ll d_\mathcal{B}$ the quantity $\epsilon_{G,\varkappa}^*$ reduces to the structure factor, $\epsilon_{G,\varkappa}^* \propto \sum_i \exp(\rmi G z_i)$, where $z_i$ are the positions of the layers $\mathcal{A}$ inside the unit cell. 
The light with the wave vector $q$ close to some reciprocal lattice vector $G/2$ exhibits  Bragg diffraction, leading to the formation of a stop band at the corresponding frequency. In the spectral vicinity of this stop band we can use the two-wave approximation~\cite{poddubny2009}
valid in the regime of relatively weak spatial modulation of the dielectric function $\varepsilon_\varkappa (z)$, which means either $d_{\mathcal A}\ll d_{\mathcal B,n}$ or $|\varepsilon_{\mathcal A}-\varepsilon_{\mathcal B}|\ll\varepsilon_{\mathcal A},\varepsilon_{\mathcal B}$,  so that $|\epsilon_{G,\varkappa}| \ll 1$.
 We take into account in Eq.~\eqref{Max} only the two harmonics, $E = (E_{K} , E_{K-G})$, with the wave vectors $K\approx q$ and $K-G \approx -q$, coupled by the  structure factor component $\epsilon_{G,\varkappa}$, and neglect all other harmonics. 
This yields the equation 
\begin{align}\label{sys2}
\bm d_{K,\varkappa} \text{$\cdot$} \bm \sigma \,  E  =[  q^2 - K^2 + G( K-G/2 )] E \:,
\end{align}
where 
\begin{align}
\bm d_{K,\varkappa} = [-q^2 \Re \epsilon_{G,\varkappa}^*,\, -q^2 \Im \epsilon_{G,\varkappa}^*,\, G(K-G/2)]\,,
\end{align}
and $\bm \sigma$ is the vector of Pauli matrices.
Introducing the dimensionless energy $\mathcal{E}=4q/G-2 \ll 1$ and the wave vector $p=4K/G-2 \ll 1$, Eq.~\eqref{sys2} can be simplified to a Dirac-like form,
\begin{equation}\label{dirac}
\begin{bmatrix}
p & -\epsilon_{G,\varkappa} \\ -\epsilon_{G}^* & -p
\end{bmatrix} \begin{bmatrix} E_{K} \\ E_{K-G} \end{bmatrix}
= \mathcal{E} \begin{bmatrix} E_{K} \\ E_{K-G} \end{bmatrix} \:,
\end{equation}
where the structure factor $\epsilon_{G,\varkappa}$ plays the role of mass. 
Equation~\eqref{dirac} yields the dispersion relation $\mathcal{E}^2 = p^2 + |\epsilon_{G,\varkappa}|^2$, describing the presence of the stop band of the width $cG|\epsilon_{G,\varkappa}|/(2\sqrt{\varepsilon_0})$ centered at the frequency $cG/(2\sqrt{\varepsilon_0})$.

The topological properties of the problem are hidden in the dependencies on the  modulation phase $\varkappa$. Particularly, when $\varkappa$ changes from $0$ to $2\pi$, the system characteristics continuously vary and for $\varkappa=2\pi$ they are the same as for $\varkappa=0$. Hence, the phase $\varkappa$ can be treated as a wave vector
in an auxiliary direction. This provides the correspondence between the  1D problem and the 2D one~\cite{Lang2012,Kraus2012b}, where the Bloch eigenstates depend on two wave vectors, $K$ and $\varkappa$. Hence, we  introduce the  Chern numbers of the allowed bands in  a standard way as
$\int_{0}^{2\pi}d\varkappa   \int_{-\pi/D}^{\pi/D}  dK \left( \partial_K A_\varkappa - \partial_{\varkappa} A_K \right)/(2\pi\i)\:,
$ where $A_K = \int \rmd z\, \sqrt{\varepsilon_\varkappa(z)}\, E^*(K,\varkappa)\partial_K \sqrt{\varepsilon_\varkappa(z)} E(K,\varkappa)$, $E(K,\varkappa)$ is the normalized solution of the wave equation~\eqref{Max} with the wave vector $K$, and $A_\varkappa$ is obtained replacing $\partial_K$ by $\partial_\varkappa$.
Using the Dirac approximation~\eqref{dirac}, valid in the vicinity of the stop band, we calculate this stop band contributions  to the Chern number of the above- and the below-laying allowed bands. The results differ only by sign and are given by the solid angle swept out by $\bm d_{K,\varkappa}$ when $K$ and $\varkappa$ are varied divided by $4\pi$~\cite{Bernevig_book}.
This angle equals to the winding number of $\epsilon_{G}^*$, i.e., the divided by $2\pi$ phase that $\epsilon_{G}^*$ gains when $\varkappa$ is changed from $0$ to $2\pi$.
Thus, the allowed band is characterized by a Chern number given by the winding number of the structure factor in above-lying stop band minus that in the below-lying stop band.  Conversely, the winding number of the structure factor  is a topological invariant of the stop band. Indeed, this number can not be changed without $\epsilon_{G,\varkappa}^*$ turning to zero for some $\varkappa$, which would mean eliminating the considered stop band.

Now we show that the winding number of the structure factor $\epsilon_{G,\varkappa}^*$ coincides with that of the reflection coefficient $r_{\infty}$ from the semi-infinite structure for the incident light frequency lying inside the corresponding stop band. 
To this end the field inside the structure is expanded as a sum of Bloch waves and the field outside $(z<0)$ as a sum of incoming $(\propto \e^{\rmi qz})$ and reflected $(\propto \e^{-\rmi qz})$ waves. We apply the boundary conditions at the interface $z=0$ with the surrounding medium with the dielectric constant $\varepsilon_{0}$ and at $z\to \infty$. In the two-wave approximation Eq.~\eqref{dirac} the resulting reflection coefficient is  equal  to  the ratio of left- and right-going waves in the corresponding eigenstate of Eq.~\eqref{dirac},
$ r_\infty (\omega)=E_{K-G}/E_{K} $~\cite{Kagan1999,Ivchenko2013},
where the wave vector $K$ should correspond to the spatially decaying eigenstate with  $p=\rmi\sqrt{|\epsilon_{G,\varkappa}|^2-\mathcal E^2}$. 
This relation directly links the measurable quantity, reflection coefficient, to the  topological properties of the Bloch states in the bulk.
Using the Hamiltonian~\eqref{dirac}  we can obtain the reflection coefficient
\begin{equation}\label{rinf}
r_\infty (\omega) =  - \frac{\epsilon_{G}^*}{\mathcal E + \rmi\sqrt{|\epsilon_{G,\varkappa}|^2-\mathcal E^2}} \:.
\end{equation}
We see that for the energy $\mathcal E$ lying inside the stop band we have full reflection, $|r_\infty|=1$, and the phase of the reflection coefficient is determined by the phase of structure factor $\epsilon_{G,\varkappa}^*$. Thus, the winding number of structure factor is equal to that of the reflection coefficient, and they both can be used to calculate the Chern numbers. 

\begin{figure*}[t]
\includegraphics[width=0.99\textwidth]{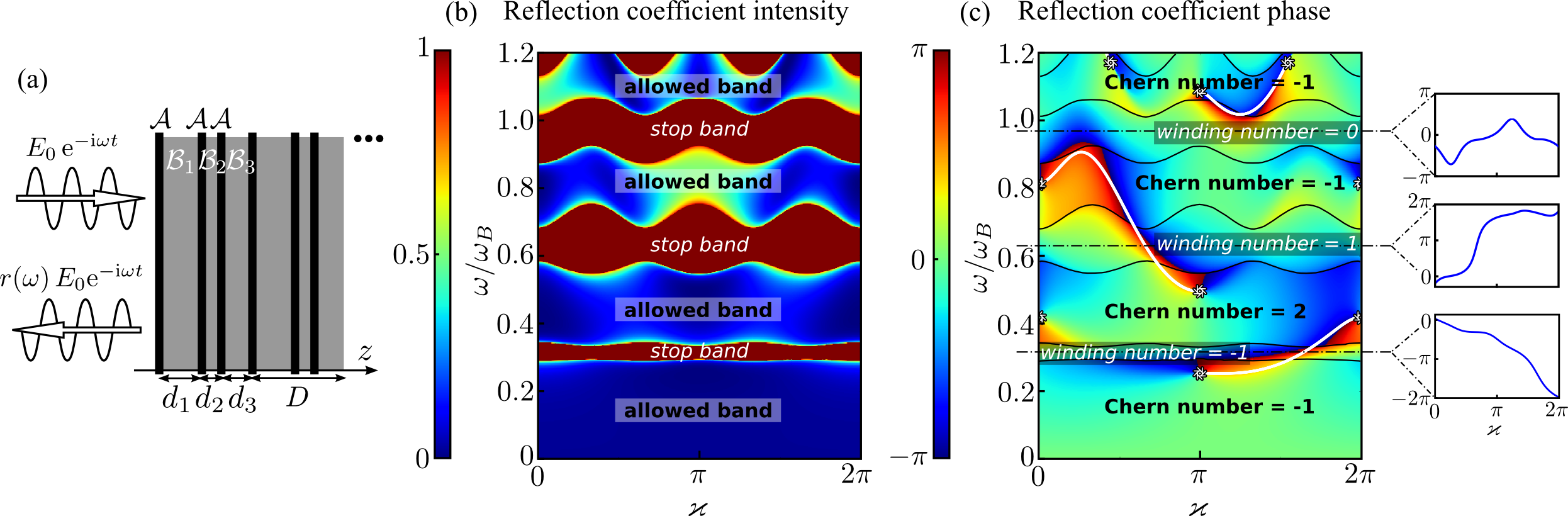}
\caption{The color map of the squared amplitude (b) and the phase (c) of the reflection coefficient from the left edge of semi-infinite 1D photonic crystal (a) as a function of light frequency $\omega$ and the parameter $\varkappa$. Solid curves show the real part of the left-edge state frequency. The subplots in panel (c) show the cross sections of the phase map at different frequencies.
 The calculation parameters are $b=1/3$, $\eta=0.5$, $n_{\mathcal A}/n_{\mathcal B}=2$, $d_{\mathcal A}/d_{\mathcal B}=0.2$.}
\label{fig1}
\end{figure*}
%

\section{Phase spectroscopy of topological states}
We have shown that from the phase of the reflection coefficient one can deduce the winding numbers of stop zones and, thus, the Chern numbers of allowed bands. 
Figure~\ref{fig1} illustrates an application of this technique to the semi-infinite 1D photonic crystal illuminated from the left.
Panel~(b) presents the absolute value of reflectivity, while panel~(c) shows the phase of the amplitude reflection coefficient  $r_\infty (\omega)$. Following the theory above, we characterize each stop band by a winding number i.e. the extra phase divided by $2\pi$ that the reflection coefficient  $r_\infty (\omega)$ gains when the parameter $\varkappa$ is varied from $0$ to $2\pi$ while $\omega$ remains inside the stop band.
If we measure the phase in some interval (e.g. from $-\pi$ to $\pi$, as in Fig.~\ref{fig1}) the winding number is equal to the number of phase cuts that cross the considered band gap, taking into account their sign. Each cut connects two phase branching points. These phase branching points are the zeros of the reflection coefficient and can be located only in the allowed bands, since in the stop band $|r(\omega)|=1$. Unless the gap closes, its winding number is invariant as there exists no way to eliminate a cut but annihilating the corresponding branching points that is impossible while they are separated by the stop band.
Comparing this result with the definition of the winding numbers through the phase of reflection spectra we find that the Chern number of a band equals to the number of phase branching points (zeros) of the reflection coefficient that are located inside this allowed band, taking into account their sign.

Now we examine the bulk-boundary correspondence, i.e. we prove  that the stop-band possesses edge states for any $\varkappa$, provided that the corresponding  winding number is nonzero. 
To this end, we 
analyze the analytical behavior of the winding number 
\begin{equation}\label{windn}
w(\omega+\i \gamma) = \frac{1}{2\pi\rmi} \int\limits_{0}^{2\pi} \frac{\partial\ln[r_{\infty}(\omega+\i\gamma,\varkappa)]}{\partial \varkappa}  \rmd\varkappa\:.
\end{equation}
in  the complex plane of frequencies. Here, we have expressed the winding number via the reflection coefficient from the semi-infinite structure and  introduced the imaginary part of the  complex frequency $\gamma$. At the first stage of the proof  we demonstrate that for every $\omega$ lying inside the stop-band there exist some $\gamma>0$ and $\varkappa$ such that 
$r_{\infty}(\omega+\i\gamma,\varkappa)=0$. Indeed, at $\gamma=0$  by definition $w(\omega) \ne 0$. On the other hand, for $\gamma\to +\infty$ the phase factors $\exp [(\rmi\omega-\gamma)d_{\mathcal A,B}\sqrt{\varepsilon_{A,B}}/c] $, describing the propagation through the layers, are quenched and only the infinitely small front part of the structure contributes to the reflection coefficient. This part is $\varkappa$-independent so the winding number vanishes.  Since  $w(\omega)\ne 0$, $\lim_{\gamma\to\infty}w(\omega+\i\gamma)=0$, and
the winding number is integer,  we conclude that there exists a jump of the winding number at certain $\gamma$. Such a discontinuity can only be caused by the singularity of the integrand in Eq.~\eqref{windn}, that can be either a zero or a pole of the reflection coefficient $r_{\infty}(\omega+\rmi\gamma,\varkappa)$. The poles of the reflection coefficient for $\gamma > 0$ are forbidden by the causality principle so the singularity corresponds to  $r_{\infty}(\omega+\rmi\gamma,\varkappa) = 0$. 

To finalize the proof  we use the identity $|r_{\infty}(\omega)|^2=1$ valid for the real values of $\omega$ lying inside the stop band. Being analytically continued onto the entire complex plane it turns into $r_{\infty}(\omega+\rmi\gamma)r_{\infty}^*(\omega-\rmi\gamma)=1$. Hence, the zero of the reflection coefficient at $\omega+\rmi\gamma$ enforces the pole at $\omega-\rmi\gamma$. Such a pole means the presence of the radiative edge state with the frequency $\omega$ and decay rate $\gamma$~\cite{Poshakinskiy2014prl}.
 The real part of the left-edge state frequency is shown in Fig.~\ref{fig1}c by solid curves.
The points where the  left-edge state appears or disappears as $\varkappa$ is varied match the branching points of reflection coefficient phase.

All the above conclusions remain valid if we introduce some coating layers at the border of the photonic crystal with vacuum. Indeed, since the coating cannot close the stop band, the corresponding winding number remains unaffected and can be still used for calculation of Chern numbers and determination of presence of edge states. This reflects the topological protection of the considered states.


%
\begin{figure*}[t]
\includegraphics[width=0.9\textwidth]{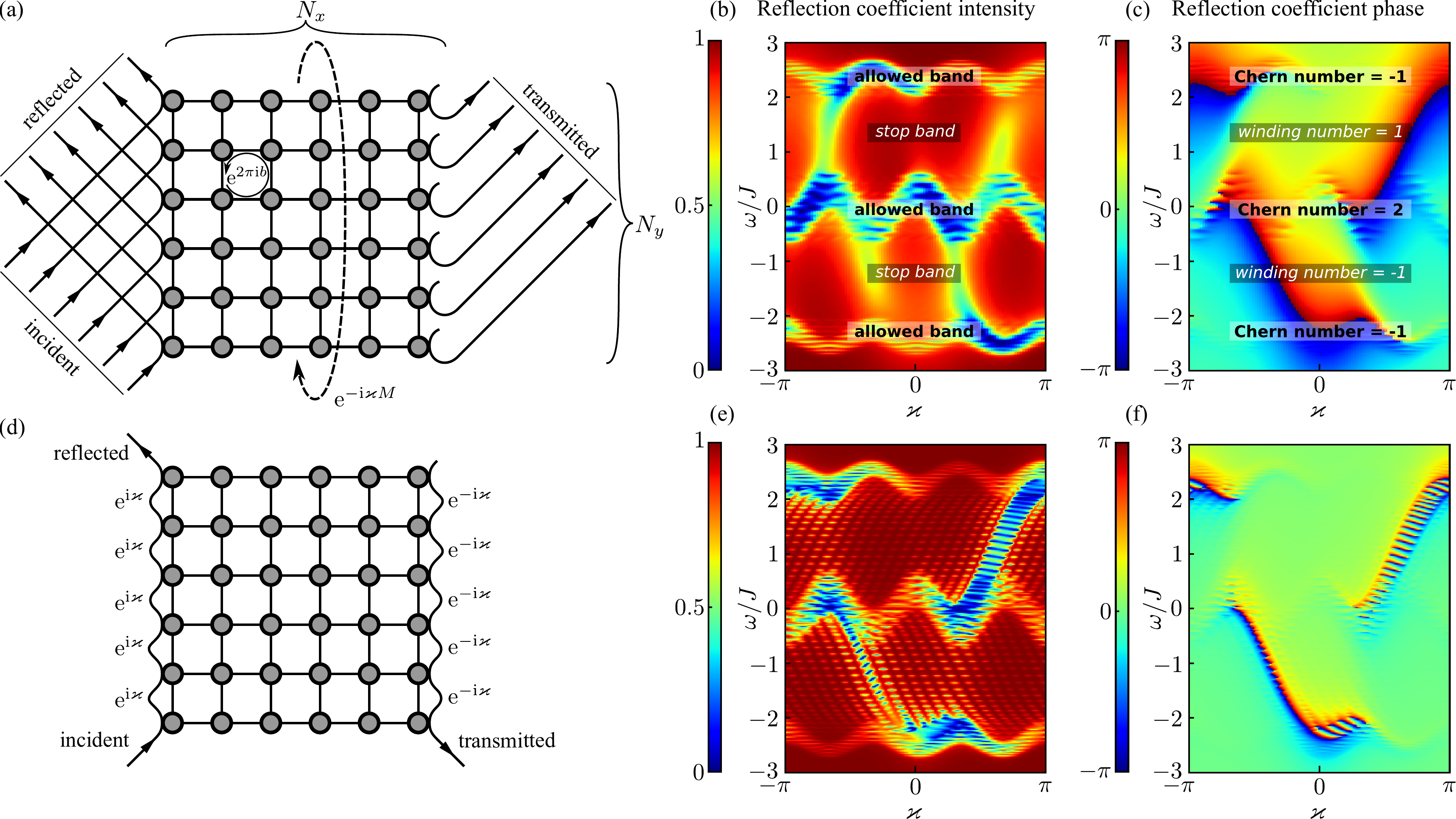}
\caption{Panels (a) and (d) show two proposed schemes of measuring of topological invariants.  Panels (b) and (c) show the calculated color map of the squared amplitude and the phase of the reflection coefficient in the multi-port scheme for the rectangular  $N_x \times N_y = 15 \times 15$ lattice,  with $b=1/3$, $\kappa/J=1$, $\gamma/J=0.01$.  Panels (e) and (f) show the squared amplitude and the phase of the reflection coefficient in the single port scheme for the same parameters except for $\kappa/J=0.01$ and  $N_x \times N_y = 21 \times 21$.}
\label{fig2}
\end{figure*}
%

\section{Application to a 2D system}

Finally, we show how the phase spectroscopy can be used to measure the topological invariants in 2D systems. As described above, in the 1D Aubry-Andr\'e lattice the topological properties follow from the dependence of the reflection coefficient phase   on the modulation parameter $\varkappa$. Now we demonstrate that in a 2D system the same topological invariants can be obtained by measuring the phase of the reflection coefficient from the edge of the system as a function of the wave vector along the edge.

We consider a 2D square optical lattice with one site per  unit cell, see Fig.~2a.
 Its realization  using the ring resonators  linked by waveguides was studied theoretically and experimentally in Refs.~\onlinecite{Hafezi2011,Hafezi2013}.
   The nontrivial topology is induced by the synthetic magnetic field that stems from the specially engineered asymmetry of link waveguides that couple the neighboring sites. The direction of the magnetic field is opposite for the clockwise and counter-clockwise resonator modes. For adiabatic links  these modes are uncoupled, so in what follows we consider only one of them. The  tight-binding Hamiltonian for the given mode reads
\begin{align}
\hat H  =  -J \sum_{n,m}\left( {\hat a}_{n,m}^{\dagger}{\hat a}_{n-1,m}+ {\hat a}_{n,m}^{\dagger}{\hat a}_{n,m-1}\e^{2\pi \rmi  b n}\right)+ {\rm H.c.} ,\nonumber
\end{align}
where $J$ is the coupling constant, ${\hat a}_{n,m}$ and ${\hat a}_{n,m}^{\dagger}$ are the photon annihilation and creation operators at the site $(n,m)$, and $2\pi b$ is the phase acquired when hopping around the unit cell. We have chosen the gauge in such a way that for rational $b=P/Q$ the magnetic unit cell has the shape $Q\times1$.

Figure~\ref{fig2}a shows the proposed scheme of measuring the topological numbers. We couple the probing waveguides to the  sites on the left and right edges of a rectangular sample. The input waveguides on the left edge are all simultaneously coherently excited at the frequency $\omega$ in such a way that the phase difference between the adjacent sites is equal to $\varkappa$. As shown in the Fig.~\ref{fig2}a, this can be easily realized by using the same light source for all input waveguides and gradually changing  the length of the input waveguide from site to site. Evidently, such an excitation scheme simulates an incident oblique  plane wave  with the wave vector $\varkappa$ along the structure edge. The reflection and transmission signals are collected from the output waveguides at the left and the right edges of the structure, respectively. The reflected and transmitted waves are given by the sum of the field over all the output waveguides at the corresponding edge, with the phases corresponding to the wave vector $\varkappa$ along the edge, which can be again realized by a proper tuning of the length of output waveguides. 

In the linear regime where $\langle \hat a_{n,m} \rangle = a_{n,m}$ the system is described by the equation set
\begin{align}\label{Liou}
-\rmi \omega {{a}}_{n,m}=&-\rmi H_{n,m;n'm'} {a}_{n',m'}\nonumber \\
-&(\kappa_{n,m}+\gamma) {a}_{n,m} 
-\sqrt{2\kappa_{n,m}} E_{n,m} ,
\end{align}
where $H_{n,m;n',m'} = \langle 0|\hat a_{n,m} \hat H \hat a^\dag_{n',m'}|0 \rangle$, $\kappa_{n,m}=\kappa (\delta_{n,1}+\delta_{n,N_x})$ describes the light coupling to the probing waveguides, i.e. $\kappa_{n,m}=\kappa$ if a probing waveguide is connected to the site $(n,m)$ and $\kappa_{n,m}=0$ if not, $E_{n,m}  =  \delta_{n,1} \e^{\rmi \varkappa m} $ is an incident field in the input waveguides on the left edge of the structure, that corresponds to the wave vector $\varkappa$ along the vertical edge,
and $\gamma$ takes into account the on-site losses.
The reflection  
coefficient for the discussed geometry of detection reads
\begin{align}
&r(\varkappa)=\sum_{m=1}^{N_y}(1+ \sqrt{2\kappa}a_{1,m} \e^{-\rmi \varkappa m}) /N_y\,.
\end{align}
 The reflection coefficient intensity  as a function of frequency and wave vector along the edge $\varkappa$ is shown in Fig.~\ref{fig2}b. One can distinguish two band gaps, where the reflectivity is close to unity. Fig.~\ref{fig2}c shows the phase of the reflection coefficient. Clearly, the winding numbers of the band gaps and the Chern numbers of the allowed bands follow from the phase map in Fig.~\ref{fig2}c in the exactly same way as for the 1D system.
 
Formal correspondence with the 1D  Aubry-Andr\'e case is attained when the twist boundary conditions are introduced on the upper and lower edges of the system.
Particularly, we roll the rectangular lattice into a cylinder as shown by the dashed arrow in Fig.~\ref{fig2}a. The additional links with the twist phase $-\varkappa N_y$, describing the flux of an effective magnetic field through the cylinder, are inserted between the corresponding sites on the upper and lower edges. These twist links are described the extra terms $\sum_{n=1}^{N_x} a^\dag_{n,1} a_{n,N_y} \e^{-\rmi \varkappa N_y} + {\rm H.c.} $ in the Hamiltonian. Then the Fourier transform along the $y$ axis shows that the only component excited is that with the wave vector $\varkappa$. This reduces the 2D system to a 1D one, similar to one described in the first part of the work.  Importantly,  for large enough $N_y$  the boundary conditions do not affect the  result of the measurement. Thus we can omit the twist links and consider the unrolled rectangle geometry, as it was done for the calculation shown in Fig.~\ref{fig2}b,c.

The multi-port scheme shown in Fig.~\ref{fig2}a can be transformed into a scheme with single input and output (see Fig.~\ref{fig2}d), which is expected to be easier in realization. The phase $\varkappa$ that corresponds to the wave vector along the edge is now introduced by tuning the length of the input and output waveguides between the neighboring lattice sites in such a way that it corresponds to the phase delay $\varkappa$ for the input waveguide and $-\varkappa$ for the output waveguide. The scheme is still described by Eq.~\eqref{Liou}, where $E_{n,m}= \delta_{n,1} E_m + \delta_{n,N_x} E'_m$ and
\begin{align}
&E_m = E_{m-1} \e^{\rmi\varkappa} + \sqrt{2\kappa}a_{1,m}, \quad (1\leq m \leq N_y)\,, \\
&E'_{m-1} = E'_{m} \e^{-\rmi\varkappa} + \sqrt{2\kappa}a_{1,m-1}, \quad (1\leq m \leq N_y)\,, \\
&E_0=1\,, \quad E'_M=0\,.
\end{align}
The reflection and transmission coefficients are equal to the fields $E_{N_y}$ and $E'_0$, respectively.
In the linear by $\kappa$ regime one can neglect the multiple tunneling of the light between the input/output waveguide and the lattice. Then the scheme presented in Fig.~\ref{fig2}d  reduces to that shown in Fig.~\ref{fig2}a. The squared amplitude and the phase of the reflection coefficient in the single input/output configuration are shown in Figs~\ref{fig2}e and f. One can see that the reflection phase in this configuration is similar to that shown in Fig.~\ref{fig2}c and  can be also  used to determine the topological indexes.

\section{Summary}

To summarize, we have shown that the Chern number of a photonic structure can be deduced from the winding number of the reflection  phase. We have demonstrated that the nonzero winding number in a certain stop band gives rise to the topological edge states. In the 1D Aubry-Andr\'e-Harper photonic crystal the winding number of the reflection coefficient in the stop band is equal to that of the structure factor characterizing the strength of the corresponding Bragg diffraction peak. To determine topological numbers of a 2D system one can use the reflection phase measured in the single- or multi-port configurations corresponding to oblique excitation. 

The authors acknowledge fruitful discussions with E.L.~Ivchenko and S. Ganeshan. This work was supported by the RFBR,  RF President Grant No. MK-6029.2014.2 and the ``Dynasty'' Foundation. M. Hafezi acknowledges the support of NSF PFC at the Joint Quantum Institute, MURI-ARO and AFOSR.

%


\end{document}